\documentclass[11pt]{amsart}
\usepackage{geometry}                
\usepackage{cite}

\usepackage{tikz-qtree}

\usepackage{url}
\usepackage{graphicx}
\usepackage{amssymb}
\RequirePackage{ifthen}
\usepackage{amsmath}
\usepackage{amsthm}
\usepackage{xcolor}
\usepackage{epstopdf}
\newtheorem{theorem}{Theorem}

\newenvironment{talk}[4][]
{\noindent\parbox{\textwidth}{%
\begin{center}%
\textbf{\boldmath #3}\\[\smallskipamount]\textsc{#2}%
\end{center}}\par\nopagebreak\bigskip\nopagebreak}%
{\bigskip\bigskip\goodbreak}

\newcommand{\nofbox}[1]{#1}

\newcommand{\phinotpsi}{\blue{\phi}}

\newcommand{\twoM}{\blue{\mathcal{S}}}

\newcommand{\mcO}{{\mycal O}}

\definecolor{applegreen}{rgb}{0.55, 0.71, 0.0}
\definecolor{armygreen}{rgb}{0.29, 0.33, 0.13}

\definecolor{caribbeangreen}{rgb}{0.0, 0.8, 0.6}

\usepackage{tikz}

\newcommand{\blue}[1]{{\color{blue}{#1}}}

\DeclareFontFamily{OT1}{rsfs}{}
\DeclareFontShape{OT1}{rsfs}{m}{n}{ <-7> rsfs5 <7-10> rsfs7 <10-> rsfs10}{}
\DeclareMathAlphabet{\mycal}{OT1}{rsfs}{m}{n}

\newcounter{mnotecount}[section]

\newcommand{\red}[1]{{\color{red}#1}}

\newcommand{\T}{\mathbb{T}}

\newcommand{\eel}[1]{\label{#1}\end{equation}}
\newcommand{\eeal}[1]{\label{#1}\end{eqnarray}}

\newcommand{\bel}[1]{\begin{equation}\label{#1}}
\newcommand{\bea}{\begin{eqnarray}}
\newcommand{\bean}{\begin{eqnarray}\nonumber}
\newcommand{\beal}[1]{\begin{eqnarray}\label{#1}}
\newcommand{\eea}{\end{eqnarray}}

\newcommand{\be}{\begin{equation}}
\newcommand{\eeq}{\end{equation}}
\newcommand{\ee}{\end{equation}}
\newcommand{\beqa}{\begin{eqnarray}}
\newcommand{\eeqa}{\end{eqnarray}}
\newcommand{\beqan}{\begin{eqnarray*}}
\newcommand{\eeqan}{\end{eqnarray*}}
\newcommand{\ba}{\begin{array}}
\newcommand{\ea}{\end{array}}

\newcommand{\mnote}[1]
{\protect{\stepcounter{mnotecount}}$^{\mbox{\footnotesize
$
\bullet$\themnotecount}}$ \marginpar{
\raggedright\tiny\em
$\!\!\!\!\!\!\,\bullet$\themnotecount: #1} }

\newcommand{\warn}[1]
{\protect{\stepcounter{mnotecount}}$^{\mbox{\footnotesize
$
\bullet$\themnotecount}}$ \marginpar{
\raggedright\tiny\em
$\!\!\!\!\!\!\,\bullet$\themnotecount: {\bf Warning:} #1} }

\newcommand{\R}{\mathbb R}
\newcommand{\N}{\mathbb N}

\newcommand{\beaa}{\begin{eqnarray*}}
\newcommand{\eeaa}{\end{eqnarray*}}

\def\ben{\begin{equation}}
\def\een{\end{equation}}
\def\bena{\begin{eqnarray}}
\def\eena{\end{eqnarray}}

\def\f(#1/#2){\frac{#1}{#2}}
\def\Frac(#1/#2){\left(\frac{#1}{#2}\right)}
\def\chris(#1-#2-#3){{\mit \Gamma}^{#1}{}_{{#2}{#3}} }
\def\tilchris(#1-#2-#3){\tilde{{\mit \Gamma}}^{#1}{}_{{#2}{#3}}}
\def\hatchris(#1-#2-#3){\hat{{\mit \Gamma}}^{#1}{}_{{#2}{#3}}}

\DeclareFontFamily{OT1}{rsfs}{}
\DeclareFontShape{OT1}{rsfs}{m}{n}{ <-7> rsfs5 <7-10> rsfs7 <10-> rsfs10}{}

\begin{document}


\renewcommand{\red}[1]{#1}
\renewcommand{\blue}[1]{#1}

\begin{talk}{Piotr T. Chru\'sciel}
{Quo Vadis, Mathematical GR?}
{Chru\'sciel, Piotr T.}

\noindent

The last five years proved that Mathematical General Relativity is more lively than ever. It got itself a Nobel Prize, with  help from Roger Penrose. Some long standing major research problems, namely stability of slowly rotating Kerr black holes and positivity of mass in all dimensions, have been announced as being solved~\cite{SchoenYau2017,Lohkamp2,SzeftelTalk}. In my talk in Oberwolfach I reported on some progress in the field, as biased by my research interests, ignoring topics which I expected to be covered by other speakers.

The topics I discussed will be split into thematic sections, no order of importance implied.

\section{Lorentzian geometry}

One of the fundamental notions in Lorentzian geometry is that of \emph{global hyperbolicity.} The standard definition proceeds as follows:  A spacetime $(M,g)$ is said to be globally hyperbolic if it is strongly causal and if for any pair of points $p,q\in M$ the \emph{causal diamond} $J^+(p)\cap J^-(q)$ is compact, if not empty. Recall that \emph{strong causality} is defined as the requirement that for every point $p$ and for every neighborhood $\mcO$ of $p$  there exists a neighborhood $\mycal U\subset \mcO$ of $p$ such that every inextendible causal curve in $M$ intersects $\mycal U$ in a connected interval.

Lorentzian geometers have been using these notions for more than 60 years by now without realising, as pointed-out by Hounnonpke and Minguzzi~\cite{HounnonkpeMinguzzi}, that for non-compact $n$-dimensional spacetimes with $n\ge 3$, global hyperbolicity is the same as requiring \emph{compactness of causal diamonds.} This is a dramatic simplification of the notion both at a conceptual level, and in applications.

Further noteworthy developments include proofs of incompleteness theorems with weaker hypotheses on the metric, see Melanie Graf's contribution to this volume.

A milestone in the understanding of the global structure of Lorentzian manifolds was the paper by Sbierski~\cite{SbierskiSchwarzschild}, who showed that the Kruskal-Szekeres manifold is inextendible in the class of $C^0$ metrics. The result was nicely complemented by the joint paper of Galloway, Ling and Sbierski~\cite{GallowayLingSbierski}, who show that globally hyperbolic timelike geodesically complete spacetimes are $C^0$ inextendible. One is then left to wonder about $C^0$-extendibility of several physically significant spacetimes: Friedman-Lemaitre-Robertson-Walker FLRW metrics? Kerr metric? Partial results have been obtained, but the arguments used for Schwarzschild do not adapt in any obvious way to these metrics. As a step towards an answer, Sbierski~\cite{SbierskiHolonomy} introduced a new technique based on holonomy to show that FLRW metrics are $C^{0,1}$--indextendible. While the current FLRW-result is not as elegant as the $C^0$-indextendibility of Schwarzschild, it should be kept in mind that being Lipschitz is the borderline regularity condition under which many things go wrong with the geometry of Lorentzian metrics, or with associated wave equations, and therefore reference~\cite{SbierskiHolonomy} provides highly relevant information.

The interesting questions of $C^0$-inextendibility of the FLRW solutions, or of Kerr, or of negative-mass Schwarzschild, remain open.

\section{Mass}
Positive-energy theorems are widely recognised as one of the most remarkable achievements of mathematical general relativity. They have found applications in the analysis of the Yamabe problem, or in proofs of uniqueness of asymptotically flat black holes. Several new  positivity proofs for asymptotically flat initial data sets have been discovered or rediscovered in recent years. Here I would only like to mention the elegant  Green function analysis by
Agostiniani, Mazzieri  and Oronzio \cite{MazzieriPET}; further approaches are described in other contributions to this volume.

Some progress has also been made on the understanding of mass for spacetimes with negative cosmological constant. In this context it is usual to consider spacetimes which have a conformal boundary at spacelike infinity \emph{\`a la Penrose}.
Quite generally, the mass of a metric is defined relatively to a background metric with a Killing vector which is timelike near the conformal boundary at infinity. Within the conformally compactifiable category, the simplest case arises for metrics which asymptote to metrics of the form
%
\begin{equation}\label{26XI21.1}
  g = V^{-2} dr^2 + r^2 h_k
  \,,
  \qquad
  V^2 = r^2 + k
  \,,
  \qquad
  k\in\{0,\pm 1\}
   \,,
\end{equation}
where $h_k $ is an Einstein metric on an $(n-1)$ dimensional manifold $N$
with scalar curvature
$$
 R(h_k)= k (n-1)(n-2)
 \,.
$$
I will refer to such metrics as \emph{asymptotically Birmingham-Kottler (BK) metrics.} There is a whole zoo of such metrics, differing by existence of  boundaries or lack thereof, and by the topology of the conformal boundary at infinity $(N,h_k)$. Our current knowledge of the sign of the mass for such metrics \cite{ChDelayHPETv1,ChGallowayHPET,ChHerzlich,Wang,BCHMN,CDW,PedersenEH,LeeNeves}
 is summarised in Table~\ref{T28VIII21.2}. The table makes it clear that a considerable amount of work remains to obtain a complete picture. 
\tikzset{baseline,every tree node/.style={align=center,anchor=north}}

\begin{table}
  \begin{tikzpicture}[scale=0.82,level 1/.style={sibling distance=-35, level distance=40},
    level 2/.style={level distance=40}]
  \Tree[.{Asymptotically Birmingham-Kottler metrics; \red{$m_{\mathrm{crit}}<0$}}
  [.{\fbox{canonical spherical}}
    [.{no bdry}
        {\nofbox{$\ge 0$ \cite{ChDelayHPETv1}}}
     ]
     [.bdry
        {\nofbox{$\ge 0$ \cite{ChGallowayHPET}}}
     ]
  ]
  [.{\qquad \fbox{Ricci flat conf.\ infinity}
     }
  [[ [.{good spin \cite{Wang,ChHerzlich}}
      [.{no bdry} {$\ge 0$}
      ]
      [.bdry
          [.$\ge 0$ ]
      ]
    ]
    [.{otherwise}
      [.{no bdry}
          {\nofbox{\red{$\ge m_{\mathrm{crit}}$ ? \cite{BCHMN}}}}
      ]
      [.bdry
          {\nofbox{$\, \exists \   m \le 0$ ?}}
      ]
    ]]]
 ]
[.{\fbox{other conf.\ infinity}}
   [.{no bdry}
      [.{ $\exists \ m< 0$ \cite{CDW,PedersenEH}}
        [.{\nofbox{\red{$\ge m_{\mathrm{crit}}$ ??}}}
        ]
      ]
    ]
    [.bdry  [.{$\mu<0$}
        [.{\nofbox{{$\ge m_{\mathrm{crit}}$ \cite{LeeNeves}}}}
        ]
      ]
       [.{otherwise}
        {\nofbox{\red{$\ge m_{\mathrm{crit}}$ ??}}}
      ]
   ]
]
]
  \end{tikzpicture}
  
  \bigskip
  
     \caption{\small Mass inequalities for asymptotically Birmingham-Kottler metrics. A double question mark indicates that no results are available; a single one indicates existence of partial results. The shorthand ``bdry'' refers to a black-hole boundary. ``Good spin'' denotes a topology where the manifold is spin \emph{and} the spin structure admits asymptotic Killing spinors.
        The case ``other conformal infinity'' includes higher genus topologies when the boundary is two-dimensional, but also e.g.\ quotients of spheres in higher dimensions.  
     Finally, $\mu$ is the mass aspect function. The  critical value of the mass $m_{\mathrm{crit}}$, assuming it exists, is expected to be determined by the conformal structure of the boundary at infinity.
     \label{T28VIII21.2}}
     \end{table}


One can also define mass for   Lorentzian metrics with conformal boundary at spacelike infinity which are more general than the asymptotically BK ones~\cite{deHaro:2000xn}. To the best of my knowledge there are no results on the properties of mass for such metrics.

\section{Evolution questions}

As already hinted-to,  authors of several key papers in this area will present their work during this meeting, and there is no point for me to duplicate this. So I will only mention some results here which will certainly not  be discussed in their talks.

\subsection{Local existence}
The ADM equations have always been at the heart of most numerical calculations in general relativity, as well as of many theoretical discussions. Their mathematical status was, and widely remains, rather unsatisfactory. There exists a round-about way of solving these equations: first solve the harmonically-reduced Einstein equations, then make a transformation which brings the metric into a desired ADM form.   A dramatic change in our understanding of this was brought by the work of Fournodavlos and Luk~\cite{FournodavlosLuk}, who show that there exists a directly well-posed formulation of the ADM equations in a Gaussian slicing (zero shift and lapse equal to one). The key is a trick, how to handle the trace of the extrinsic curvature of the metric.
%

\subsection{Stability of de Sitter spacetime, higher dimensional $\mycal I$}
One such result concerns the question of stability of even-dimensional de Sitter  spacetime under small vacuum perturbations. The $3+1$ dimensional case has been settled by Friedrich in a landmark paper~\cite{F1}. This has been followed by an inspired observation by  Anderson~\cite{AndersonCIE}, that the \emph{Fefferman-Graham  obstruction tensor} can be used as a tool to prove stability of higher-dimensional de Sitter spacetime with odd space-dimensions. (An alternative approach  can be found in \cite{RingstroemdS}.) In a joint paper  with  Anderson\cite{AndersonChruscielConformal} we used the same idea to construct even-dimensional spacetimes with vanishing cosmological constant and a smooth conformal completion at null infinity. An embarassing mistake in the proof of well-posedness of the associated evolution equations in \cite{AndersonChruscielConformal}, relevant both to the stability of de Sitter and to existence of null infinity, has been pointed-out and corrected in~\cite{KaminskiFG}.

\subsection{Electrically charged Robinson-Trautman solutions}

An approach to the Einstein equations which is very familiar to an Oberwolfach audience is by solving a Cauchy problem. It is somewhat surprising that the charged Robinson-Trautman metrics cannot be handled in this way: after introducing an ansatz for the metric, and solving most of the Einstein-Maxwell equations, one ends up with
a set of coupled equations, one of which is parabolic to the future, the other to the past~\cite{LunChow}. A well posed problem for the \emph{linearised} equations is obtained by introducing a spectral projection operator, with solutions determined by the projection of a set of free  data at a characteristic surface $u=u_-$, with the complementing projection of the free data provided at a characteristic surface $u=u_+>u_-$~\cite{ChTodEMRT}.
The question then arises whether one can prove something similar for the nonlinear problem.

For definiteness we note that the metric takes the form
\begin{equation}\label{5II21.PT}
ds^2=-\left(-2r\frac{P_{,u}}{P}+K-2\frac{M}{r}+\frac{Q_0^2}{r^2}\right)du^2
 -2dudr+\frac{2r^2d\zeta d\overline{\zeta}}{P^2},
\end{equation}
(quoted from \cite{LunChow}, see \cite{Exactsolutions2} 
for a derivation). Here $P(u,\zeta,\overline{\zeta})$ is the dynamical variable determining the metric of the 2-surface $\mathcal{S}$ coordinatised by a complex coordinate $\zeta$, $K=\Delta\red{\log}  P$ is the Gauss curvature of the metric
$$
 g:= 2 P^{-2} d\zeta d\overline{\zeta}
 \,,
$$
$\Delta$ is the Laplace operator of $g$,
$Q_0$ is a nonzero real number and $M(u,\zeta,\overline{\zeta})$ is a dynamical variable  which also enters in  the electromagnetic potential $A$:
\[
A=(M-\frac{Q_0}{r})du \,.
\]
The Einstein-Maxwell equations  reduce to a pair of  equations which involve derivatives up to order four:
\begin{eqnarray}
     (\red{\log} P),_u & = & -{1\over {4Q_{0}^2}} \Delta M
\,,
  \label{eq:rt1}  \\
           M,_u & = &
 {  -{3M\over {4Q_{0}^2}} \Delta M + {1\over 4}\Delta^2 \red{\log} P-
           {1\over 4 Q_{0}^2} |d M|^2_g }
\,,   \label{eq:rt2}
\end{eqnarray}
where $|\cdot|_g$ denotes the norm with respect to the metric $g$.

It turns out that the nonlinear equations with small part-initial-part-final data can be solved provided a certain functional inequality holds.
In order to present the problem, for definiteness we consider a square torus
$$
 \twoM = \T^2:=[0,2\pi]\times[0,2\pi]
$$
with the flat metric inherited from $\R^2$,  and we note that a similar approach applies to other topologies. An orthonormal basis of $L^2$ consisting of  eigenfunctions of the Laplacian is given by the collection of functions
\begin{equation}\label{16II21.1}
  f_{\vec \ell}\, (\vec x) = \frac{1}{2\pi} e^{i \vec \ell \vec x}
   \,,
    \quad
    \vec \ell \in \N^2
    \,.
\end{equation}

Let $G:\R\to\R$ be a smooth function with $G(0)=0$. We denote by
$\red{(G\circ \phinotpsi)_{\vec \ell}}$ the Fourier coefficients of  the composition $G\circ \phinotpsi $,
\begin{equation}\label{16II21.3}
  \red{G\circ \phinotpsi  } (\vec x) = \sum_{
    {\vec \ell} \in \N^2} \red{(G\circ \phinotpsi)_{\vec \ell}} \, f_{\vec \ell}\, (\vec x)
   \,,
\end{equation}
and by $\phinotpsi _{\vec \ell}$ the Fourier coefficients of $\phinotpsi $:
\begin{equation}\label{16II21.4}
   \phinotpsi  (\vec x) = \sum_{
    {\vec \ell} \in \N^2} \phinotpsi _{\vec \ell} \, f_{\vec \ell}\, (\vec x)
    \,.
\end{equation}

Consider the well known inequality: for $s>1$,
\begin{equation}\label{18II21.1zxc}
  \|G\circ \phinotpsi \|_{H^s(\twoM)} \le C(\|\phinotpsi \|_{L^\infty(\twoM)})  \| \phinotpsi \|_{H^s(\twoM)}
\,,
\end{equation}
where the strictly increasing function $C$ depends upon $G$ and $s$.
After multiplying the function $C$ by a constant if necesary, \eqref{18II21.1zxc}
is equivalent to
\begin{equation}\label{18II21.2}
   \sum_{\vec \ell\in \N^2} (1+ |\vec \ell|)^{2s}
   \red{(G\circ \phinotpsi)_{\vec \ell}}^2
 \le
 C(\|\phinotpsi \|_{L^\infty(\twoM)})
  \Big( \sum_{\vec \ell\in \N^2} (1+ |\vec \ell|)^{2s}
   \phinotpsi _{\vec \ell}^2
    \Big)
  \,.
\end{equation}

It turns out the full Einstein-Maxwell equations for  Robinson-Trautman metrics can be solved, at least in the small-data regime, if a variation  of this inequality holds for functions $\phinotpsi $   which depend upon a time parameter $t$. We formulate the desired inequality as a question:

Is it true that, given a smooth function $G$ with a first order zero at the origin and a real number $s>1$,
there exists a constant $C_2=C_2(G,s)$ such that for all $\phinotpsi (t)$ satisfying
\begin{equation}\label{18II21.5a}
   \sum_{\vec \ell\in \N^2} (1+ |\vec \ell|)^{2s}
    \red{\sup_{t\in[t_-,t_+]}} \phinotpsi _{\vec \ell}^2(t)
    \le 1
\,.
\end{equation}
we have
\begin{equation}\label{18II21.5}
     \sum_{\vec \ell\in \N^2} (1+ |\vec \ell|)^{2s}
  \red{\sup_{t\in[t_-,t_+]}} \red{(G\circ \phinotpsi)_{\vec \ell}}^2(t)
 \le
 C_2
  \sum_{\vec \ell\in \N^2} (1+ |\vec \ell|)^{2s}
    \red{\sup_{t\in[t_-,t_+]}} \phinotpsi _{\vec \ell}^2(t)
\  ?
\end{equation}

Note that a direct application of \eqref{18II21.1zxc} gives a different inequality:
\begin{eqnarray}
\lefteqn{ \red{\sup_{t\in[t_-,t_+]}}
   \ \Big( \sum_{\vec \ell\in \N^2} (1+ |\vec \ell|)^{2s}
   \red{(G\circ \phinotpsi)_{\vec \ell}}^2(t)
    \Big)
}
 &&
\\
 &&
 \le
 C(  \red{\sup_{t\in[t_-,t_+]}} \|\phinotpsi (t)\|_{L^\infty(\twoM)})
  \red{\sup_{t\in[t_-,t_+]}} \Big( \sum_{\vec \ell\in \N^2} (1+ |\vec \ell|)^{2s}
     \phinotpsi _{\vec \ell}^2(t)
    \Big)
  \,.
  \nonumber
\end{eqnarray}
Assuming that \eqref{18II21.5} holds, we have the conditional result~\cite{ChTodEMRT}:

\begin{theorem}
  \label{T1II21.1a}
Let $    s, u_\pm \in \R$ with $s>1$ and $u_-<u_+$.  If the inequality \eqref{18II21.5} holds,
there exists a unique smooth solution
of the Einstein-Maxwell Robinson-Trautman equations
defined on $[u_,u_+]\times \T^2$ with part of  sufficiently small spectral data of $(M,P)$ prescribed at $u_-$, and the remaining part on $u_+$.
\end{theorem}

The question then arises, whether \eqref{18II21.5} holds true.

\section{Interferometers}

Yet another recent Nobel prize for general relativity was awarded for the first direct observation of gravitational waves. The detection involved a Michelson interferometer, which works as follows: a laser sends light to a beam splitter, the two resulting  beams  bounce  back and forth a few times between  mirrors and interfere  at the output port. The freshman calculation is to determine the affine parameter needed for the roundtrip along the geodesics connecting the (freely falling) beam splitter and mirrors. A somewhat more sophisticated approach is to calculate the leading perturbation in an eikonal expansion for the Maxwell equation. With some hand waving one recovers the geodesic approximation just described,
but one quickly realises that the solutions so obtained are ambiguous and possibly coordinate dependent. So the question arises, is there a way of describing this problem which guarantees existence of a unique solution, with an unambiguous answer for the interference pattern. Note that geometric uniqueness will also guarantee coordinate independence of the final result.

In \cite{ChMielingPalenta} we have observed that one can solve uniquely the Maxwell equations by describing the laser, and the mirrors, as a boundary-value problem for the Maxwell equations with analytic boundary data given on an infinite \emph{timelike} hyperplane in a space-time with an analytic metric
\begin{equation}\label{28VIII21.23}
  g_{\mu\nu} =
  \eta_{\mu\nu} + \epsilon A_{\mu\nu} \cos (\vec k\vec x - \omega_g t)
  \,,
\end{equation}
where $\eta_{\mu\nu}$ is the Minkowski metric and $A_{\mu\nu}$ is an $\eta$-traceless tensor with constant entries satisfying $A_{0\mu}=0=A_{ij} k^j$.
Existence of solutions, and analyticity in $\epsilon$ is guaranteed by the Cauchy-Kovalevskaya theorem. Uniqueness within the class of smooth solutions is guaranteed by a theorem of Holmgren. One can then calculate explicitly the first term in a (convergent) expansion of the solution in terms of powers of $\epsilon$, and find that is has a Laurent expansion in terms of
$\omega_g/\omega$, where $\omega_g$ is the frequency of the gravitational wave and $\omega$ is the frequency of the  Maxwell wave emitted by the laser. It turns out that this is precisely what is needed to validate the application of eikonal expansions for the problem at hand. One also shows that those ambiguities which remain do not affect the interference pattern at the level of accuracy available experimentally.

The above relies heavily on the analyticity of the metric \eqref{28VIII21.23}, which is a poor man's approximation to the physical metric. This last metric has no reason to be analytic. One is thus led to the question, whether there exists a way to justify rigorously the results of LIGO-type interferometric experiments without assuming analyticity?

\smallskip
\noindent{\sc Acknowledgements} Research supported in part by the Austrian Science Fund (FWF), Project P34274, and by the Vienna University Research Platform TURIS. Part of this research was performed while the author was visiting the Institute for Pure and Applied Mathematics (IPAM), which is supported by the National Science Foundation (Grant No. DMS- DMS-1925919).

\bibliographystyle{amsplain}\providecommand{\bysame}{\leavevmode\hbox to3em{\hrulefill}\thinspace}

\begin{thebibliography}{10}

\bibitem{MazzieriPET}
V.~{Agostiniani}, L.~{Mazzieri}, and F.~{Oronzio}, \emph{{A Green's function
  proof of the Positive Mass Theorem}},  (2021), arXiv:2108.08402 [math.DG].

\bibitem{AndersonCIE}
M.T. Anderson, \emph{Existence and stability of even dimensional asymptotically
  de {S}itter spaces},  (2004), arXiv:gr-qc/0408072.

\bibitem{AndersonChruscielConformal}
M.T. Anderson and P.T. Chru\'{s}ciel, \emph{Asymptotically simple solutions of
  the vacuum {E}instein equations in even dimensions}, Commun.\ Math.\ Phys.
  \textbf{260} (2005), 557--577, arXiv:gr-qc/0412020. \MR{MR2183957}

\bibitem{BCHMN}
H.~Barzegar, P.T. Chru\'sciel, M.~{H\"orzinger}, M.~Maliborski, and L.~Nguyen,
  \emph{{On the energy of asymptotically Horowitz-Myers metrics}}, Phys.\ Rev.\
  D \textbf{101} (2019), 024007, 16, arXiv:1907.04019 [gr-qc].

\bibitem{ChDelayHPETv1}
P.T. Chru\'{s}ciel and E.~Delay, \emph{{The hyperbolic positive energy
  theorem}},  (2019), arXiv:1901.05263v1 [math.DG].

\bibitem{CDW}
P.T. Chru\'sciel, E.~Delay, and R.~Wutte, \emph{{Hyperbolic energy and Maskit
  gluings}},  (2021), arXiv:2112.00095 [math.DG].

\bibitem{ChGallowayHPET}
P.T. Chru\'sciel and G.J. Galloway, \emph{{Positive mass theorems for
  asymptotically hyperbolic Riemannian manifolds with boundary}},  (2021),
  arXiv:2107.05603 [gr-qc].

\bibitem{ChHerzlich}
P.T. Chru\'{s}ciel and M.~Herzlich, \emph{The mass of asymptotically hyperbolic
  {R}iemannian manifolds}, Pacific Jour.\ Math. \textbf{212} (2003), 231--264,
  arXiv:math/0110035 [math.DG]. \MR{MR2038048 (2005d:53052)}

\bibitem{ChTodEMRT}
P.T. Chru\'{s}ciel and P.~Tod, \emph{On {Robinson-Trautman metrics with a
  Maxwell} field},  (2021).

\bibitem{FournodavlosLuk}
G.~Fournodavlos and J.~Luk, \emph{{Asymptotically Kasner-like singularities}},
  (2020), arXiv:2003.13591 [gr-qc].

\bibitem{F1}
H.~Friedrich, \emph{On the regular and the asymptotic characteristic initial
  value problem for {E}instein's vacuum field equations}, Proc.\ Roy.\ Soc.\
  London Ser.\ A \textbf{375} (1981), 169--184. \MR{MR618984 (82k:83002)}

\bibitem{GallowayLingSbierski}
G.J. Galloway, E.~Ling, and J.~Sbierski, \emph{Timelike completeness as an
  obstruction to {$C^0$}-extensions}, Commun.\ Math.\ Phys. \textbf{359}
  (2018), 937--949. \MR{3784536}

\bibitem{deHaro:2000xn}
{S.~de} Haro, S.N. Solodukhin, and K.~Skenderis, \emph{Holographic
  reconstruction of spacetime and renormalization in the {AdS/CFT}
  correspondence}, Commun.\ Math.\ Phys. \textbf{217} (2001), 595--622,
  arXiv:hep-th/0002230.

\bibitem{HounnonkpeMinguzzi}
R.A. Hounnonkpe and E.~Minguzzi, \emph{{Globally hyperbolic spacetimes can be
  defined without the 'causal' condition}}, Class.\ Quantum Grav. \textbf{36}
  (2019), 197001, arXiv:1908.11701 [gr-qc].

\bibitem{KaminskiFG}
W.~{Kami{\'n}ski}, \emph{{Well-posedness of the ambient metric equations and
  stability of even dimensional asymptotically de Sitter spacetimes}},  (2021),
  arXiv:2108.08085 [gr-qc].

\bibitem{LeeNeves}
D.A. Lee and A.~Neves, \emph{The {P}enrose inequality for {a}symptotically
  {l}ocally {h}yperbolic spaces with nonpositive mass}, Commun.\ Math.\ Phys.
  \textbf{339} (2015), 327--352. \MR{3370607}

\bibitem{Lohkamp2}
J.~{Lohkamp}, \emph{{The Higher Dimensional Positive Mass Theorem II}},
  (2016), arXiv:1612.07505 [math.DG].

\bibitem{LunChow}
A.W.C. Lun and E.W.M. Chow, \emph{The role of apparent horizon on the evolution
  of {R}obinson-{T}rautman {E}instein-{M}axwell spacetimes}, Confronting the
  infinite ({A}delaide, 1994), World Sci. Publ., River Edge, NJ, 9 1995,
  pp.~279--292. \MR{1418659, arXiv:gr-qc/9409024}

\bibitem{ChMielingPalenta}
T.~Mieling, P.T. Chru\'sciel, and S.~Palenta, \emph{The electromagnetic field
  in gravitational wave interferometers},  (2021), arXiv:2107.07727 [gr-qc].

\bibitem{PedersenEH}
H.~Pedersen, \emph{Eguchi-{H}anson metrics with cosmological constant},
  Classical Quantum Gravity \textbf{2} (1985), no.~4, 579--587. \MR{795103}

\bibitem{RingstroemdS}
H.~Ringstr{\"om}, \emph{{On the topology and future of stability of the
  universe}}, Oxford mathematical monographs, Oxford Univ.\ Press, Oxford,
  2013.

\bibitem{SbierskiSchwarzschild}
J.~Sbierski, \emph{{The $C^0$-inextendibility of the Schwarzschild spacetime
  and the spacelike diameter in Lorentzian Geometry}}, Jour.\ Diff.\ Geom.
  \textbf{108} (2018), 319--378, arXiv:1507.00601 [gr-qc]. \MR{3763070}

\bibitem{SbierskiHolonomy}
\bysame, \emph{{On holonomy singularities in general relativity and the
  $C^{0,1}_{\mathrm{loc}}$-inextendibility of spacetimes}},  (2020),
  arXiv:2007.12049 [gr-qc].

\bibitem{SchoenYau2017}
R.~Schoen and S.-T. Yau, \emph{{Positive Scalar Curvature and Minimal
  Hypersurface Singularities}},  (2017), arXiv:1704.05490 [math.DG].

\bibitem{Exactsolutions2}
H.~Stephani, D.~Kramer, M.~MacCallum, C.~Hoenselaers, and E.~Herlt, \emph{Exact
  solutions of {E}instein's field equations}, Cambridge Monographs on
  Mathematical Physics, Cambridge University Press, Cambridge, 2003 (2nd ed.).
  \MR{MR2003646 (2004h:83017)}

\bibitem{SzeftelTalk}
J.~Szeftel, \emph{Vienna relativity seminar}, June 2021.

\bibitem{Wang}
X.~Wang, \emph{Mass for asymptotically hyperbolic manifolds}, Jour.\ Diff.\
  Geom. \textbf{57} (2001), 273--299. \MR{MR1879228 (2003c:53044)}

\end{thebibliography}
\providecommand{\MR}{\relax\ifhmode\unskip\space\fi MR }
\providecommand{\MRhref}[2]{%
  \href{http://www.ams.org/mathscinet-getitem?mr=#1}{#2}
}
\providecommand{\href}[2]{#2}

\end{talk}
\end{document}